\documentclass[12pt]{article}
\usepackage{amsmath}
\usepackage{graphicx}
\usepackage{amsfonts}
\usepackage{amssymb}
\newtheorem{theorem}{Theorem}[section]

\numberwithin{equation}{section}

\begin{document}

\title{Non linear stability of an expanding universe with $S^{1}$ isometry group}
\author{Yvonne Choquet-Bruhat and Vincent Moncrief}
\maketitle
\begin{abstract}
We prove the existence for an infinite proper time in the expanding direction
of spacetimes satisfying the vacuum Einstein equations on a manifold of the
form $\Sigma\times S^{1}\times R$ where $\Sigma$ is a compact surface of genus
$G>1.$ The Cauchy data are supposed to be invariant with respect to the group
$S^{1}$ and sufficiently small , but we do not impose a restrictive hypothesis
made in the previous work [1].
\end{abstract}

\section{Introduction.}

An einsteinian universe is a pair $(V,^{(4)}g)$, with $V$ a smooth 4
dimensional manifold and $^{(4)}g$ a lorentzian metric on $V$ which satisfies
the Einstein equations. Such a universe satisfies the classical causality
requirements if it is globally hyperbolic, equivalently if $V$ is a product
$V=M\times R$ with each $M_{t}\equiv M\times\{t\}$ space like and a Cauchy
surface, i.e. intersected once by each causal curve (timelike or null). It is
well known that given a 3 dimensional manifold endowed with a properly
riemannian metric $\bar{g}$ and a symmetric 2 tensor $K$ satisfying the
Einstein constraint equations, there exists (modulo appropriate functional
hypotheses on the data) a globally hyperbolic vacuum (the theorem extends to
classical sources which admit of a well posed Cauchy problem) einsteinian
universe such that $M_{t_{0}}$ is a Cauchy surface, and the spacetime metric
$^{(4)}g$ induces on $M_{t_{0}}$ the metric $\bar{g}$ while $K$ is the
extrinsic curvature of $M_{t_{0}}.$ This solution is unique, up to isometry,
in the class of maximal spacetimes (i.e. which cannot be embedded in a larger
one). In spite of its formulation using $R$ this solution is a local one: $t$
is just a coordinate, it has no intrisic meaning: the physically meaningful
quantity is the proper time, determined by the metric $^{(4)}g.$ The main
problems which remain open in this field are the infinite proper time
existence, or the formation of singularities and their nature.

In this article we will prove the existence for an infinite proper time, in
the expanding direction, of vacuum einsteinian universes with Cauchy data
which are in a neighbourhood of a vacuum einsteinian universe defined as follows.

\begin{itemize}
\item $V=M\times R$ is such that $M$ is a compact manifold of the form
$M=S^{1}\times\Sigma,$ with $\Sigma$ a smooth, orientable surface.

\item  The spacetime metric is invariant under the group $S^{1}.$ It is given
by:
\begin{equation}
^{(4)}g=-4dt^{2}+2t^{2}\sigma+\theta^{2}%
\end{equation}
with $\sigma$ a metric on $\Sigma$ independent of $t$ and of scalar curvature
$-1$, and $\theta$ a 1-form on $S^{1}$. In local coordinates $x^{a}\,$on
$\Sigma$ and $dx^{3}$ on $S^{1}$ we have:
\[
\sigma=\sigma_{ab}dx^{a}dx^{b},\text{ \ \ }a,b=1,2\text{ \ \ \ }\theta
=dx^{3}.
\]
The property
\begin{equation}
R(\sigma)=-1
\end{equation}
implies, by the Gauss Bonnet theorem, that the surface $\Sigma$ has genus $G>1.$
\end{itemize}

\textbf{Remark} (important fact for the Thurston classification). The
lorentzian 3 metric $-4dt^{2}+2t^{2}\sigma$\ is homogeneous in $t$, but not
the 4 metric.

The above universe is a particular case of the ones described in the next section.

\section{$S^{1}$ invariant einsteinian universes.}

The spacetime manifold is a product $S^{1}\times\Sigma\times R$, where
$\Sigma$ is a smooth, compact, orientable 2 - manifold of genus $G>1$. The
spacetime 4 metric is invariant under the action of the group $S^{1}$, with
spacelike orbits $S^{1}\times\{x\}\times\{t\}$. We restrict here our study to
the so called \textbf{polarized case} where the orbits are orthogonal to 3
dimensional lorentzian sections. The metric can be written, without loss of
generality and for later convenience, in the form:
\begin{equation}
^{(4)}g=e^{-2\gamma(3)}g+e^{2\gamma}(\theta)^{2},
\end{equation}
with $\gamma$ a scalar function and $^{(3)}g$ an arbitrary lorentzian 3-metric
on $\Sigma\times R$ which we write under the usual form adapted to the Cauchy
2+1 splitting:
\begin{equation}
^{(3)}g=-N^{2}dt^{2}+g_{ab}(dx^{a}+\nu^{a}dt)(dx^{b}+\nu^{b}dt)
\end{equation}
$N,$ $\nu,$ $g=g_{ab}dx^{a}dx^{b}$\ are $t$ dependent scalar (lapse), vector
(shift), metric on $\Sigma$ and $\theta=dx^{3}$ a 1-form on $S^{1},x^{3}$ a
periodic coordinate.We denote by $\partial_{\alpha}$ the Pfaff derivatives in
the coframe $\theta^{0}=dt,$\ $\theta^{a}=dx^{a}+\nu^{a}dt$, greek indices
taking the values 0, 1 or 2. It holds that:
\begin{equation}
\partial_{0}=\frac{\partial}{\partial t}-\nu^{a}\frac{\partial}{\partial
x^{a}}%
\end{equation}
We introduce the extrinsic curvature $k_{t}$\ of $\Sigma_{t}$ i.e. we set
\begin{equation}
k_{ab}=-\frac{1}{2N}\hat{\partial}_{0}g_{ab},
\end{equation}
with $\hat{\partial}_{0}$ the operator on time dependent tensors on $\Sigma$
given by:
\begin{equation}
\hat{\partial}_{0}=\frac{\partial}{\partial t}-L_{\nu_{t}}.
\end{equation}
with $L_{\nu_{t}}$ the Lie derivative with respect to $\nu_{t}.$ The mean
curvature $\tau$ of $\Sigma_{t}$, which will play a fundamental role in our
later estimates, is:
\begin{equation}
\tau\equiv g^{ab}k_{ab}.
\end{equation}

\section{Equations.}

The metric $^{(4)}g$ is supposed to satisfy the vacuum Einstein equations,
\[
Ricci(^{(4)}g)=0.
\]

The equations $^{(4)}R_{\alpha3}=0$ are identically satisfied by the metric 2.1..

\subsection{Wave equation for $\gamma.$}

The equation $^{(4)}R_{33}=0$ implies that the function $\gamma$ satisfies the
wave equation on ($\Sigma\times R,^{(3)}g)$. This equation reads:
\begin{equation}
-N^{-1}\partial_{0}(N^{-1}\partial_{0}\gamma)+Ng^{ab}\nabla_{a}(N\partial
_{b}\gamma)+N^{-1}\tau\partial_{0}\gamma=0
\end{equation}

\subsection{3 dimensional Einstein equations.}

When $^{(4)}R_{3\alpha}=0$ and $^{(4)}R_{33}=0$ the equations $^{(4)}%
R_{\alpha\beta}=0$ are equivalent to Einstein's equations on $\Sigma\times R$
for the metric $^{(3)}g$ with source the stress energy tensor of the scalar
field $\gamma,$ namely:
\begin{equation}
^{(3)}R_{\alpha\beta}=2\partial_{\alpha}\gamma\partial_{\beta}\gamma
\end{equation}
In dimension 3 the Einstein equations are non dynamical, except for the
conformal class of $g.$ We set on $\Sigma\times R$%
\[
g_{ab}=e^{2\lambda}\sigma_{ab},
\]

We impose that, on each $\Sigma_{t},$ $\sigma_{t}$ has scalar curvature
$R(\sigma_{t})=-1,$ i.e. $\sigma_{t}\in M_{-1},$ which is no restriction since
any metric $\sigma$ on $\Sigma,$ which is of genus greater than 1, is
conformal to such a metric, and $e^{2\lambda}$ is to be determined.

As a gauge condition we suppose that the mean extrinsic curvature $\tau$ is
constant on each $\Sigma_{t},$ i.e. depends only on $t.$ We will construct an
expanding space time, i.e. we take $\tau$ to be negative and increasing from a
value $\tau_{0}<0$. The moment of maximum expansion will be attained if $\tau$
reaches the value $\tau=0.$ For convenience we define the parameter $t$ by
\begin{equation}
t=-\tau^{-1},\text{ \ \ \ }t\geq t_{0}.
\end{equation}
The following equations (momentum constraint) hold on $\Sigma_{t}$:%

\begin{equation}
-Ne^{-2\lambda(3)}R_{0a}\equiv D_{b}h_{a}^{b}=L_{a}\equiv-2D_{a}\gamma
\dot{\gamma}%
\end{equation}
with $h_{ab}=k_{ab}-\frac{1}{2}g_{ab}\tau$ the traceless part of $k_{t},D_{a}$
the covariant derivative in the metric $\sigma_{t}$ and indices raised with
$\sigma^{ab},$ and we have set:
\[
\dot{\gamma}=e^{2\lambda}\gamma^{\prime}\text{ with }\gamma^{\prime}%
=N^{-1}\partial_{0}\gamma.
\]
Given $\sigma,\gamma$ and $\dot{\gamma}$ this is a linear equation for $h$ on
$\Sigma_{t},$ independent of $\lambda.$ The general solution is the sum of a
solution $q$ of the homogeneous equation, a trace and divergence free tensor
called a TT tensor, and a conformal Lie derivative $r.$ Such tensors are
$L^{2}(\sigma)$ orthogonal.

The so called hamiltonian constraint on $\Sigma_{t},$ on the other hand, is
given by%

\begin{equation}
2N^{-2(3)}S_{00}\equiv R(^{(3)}g)-g^{ac}g^{bd}h_{bc}h_{ad}+\frac{1}{2}\tau
^{2}=2(N^{-2}\partial_{0}\gamma\partial_{0}\gamma+g^{ab}\partial_{a}%
\gamma\partial_{b}\gamma)
\end{equation}
\bigskip When $g_{ab}=e^{2\lambda}\sigma_{ab}$ it becomes a semilinear
elliptic equation in $\lambda:$
\begin{equation}
\Delta\lambda=f(x,\lambda)\equiv p_{1}e^{2\lambda}-p_{2}e^{-2\lambda}+p_{3},
\end{equation}
with
\[
p_{1}\equiv\frac{1}{4}\tau^{2},p_{2}\equiv\mid\overset{.}{\gamma}\mid
^{2}+\frac{1}{2}\mid h\mid^{2},p_{3}\equiv\frac{1}{2}R(\sigma)-|D\gamma|^{2}%
\]
\ 

The equation $^{(3)}R_{00}=\rho_{00}$ gives for $N$ the linear elliptic
equation
\begin{equation}
\Delta N-\alpha N=-e^{2\lambda}\partial_{t}\tau\text{ with }\alpha\equiv
e^{-2\lambda}(\mid h\mid^{2}+\mid\dot{u}\mid^{2})+\frac{1}{2}e^{2\lambda}%
\tau^{2}%
\end{equation}

The shift $\nu$ satisfies the equation (resulting from the expression for $h)$%

\begin{equation}
(L_{\sigma}n)_{ab}\equiv D_{a}n_{b}+D_{b}n_{a}-\sigma_{ab}D_{c}\nu^{c}%
=f_{ab}\text{ with }n_{a}\equiv\nu_{a}e^{-2\lambda}%
\end{equation}%
\[
f_{ab}\equiv2Ne^{-2\lambda}h_{ab}+\partial_{t}\sigma_{ab}-\frac{1}{2}%
\sigma_{ab}\sigma^{cd}\partial_{t}\sigma_{cd}%
\]

We require the metric $\sigma_{t}$ to lie in some chosen cross section
$Q\rightarrow\psi(Q)$ of the fiber bundle $M_{-1}\rightarrow T_{eich}$ with
$T_{eich}$ (Teichm\.{u}ller) the space of classes of conformally inequivalent
riemannian metrics, identified with $R^{6G-6}.$ The solvability condition for
the shift equation determines $dQ/dt$ in terms of $h.$ One obtains an ordinary
differential system for the evolution of $Q$ by requiring that the
$t-$dependent spatial tensor $^{(3)}R_{ab}-2\partial_{a}u\partial_{b}u,$ which
is TT by the previously solved equations, be $L^{2}$ orthogonal to TT tensors.

\section{Local existence theorem.}

The Cauchy data on $\Sigma_{t_{0}}$ are:

1. A $C^{\infty}$ riemannian metric $\sigma_{0}$ which projects onto a point
$Q(t_{0})$ of $T_{eich}$ and a $C^{\infty}$ tensor $q_{0}$ which is TT in the
metric $\sigma_{0}.$ The spaces $W_{s}^{p}$ and $H_{s}\equiv W_{s}^{2}$ are
the usual Sobolev spaces of tensor fields on the riemannian manifold
$(\Sigma,\sigma_{0}).$

2. Cauchy data for $\gamma$ and $\dot{\gamma}$ on $\Sigma_{0}$, i.e.
\[
\gamma(t_{0},.)=\gamma_{0}\in H_{2},\dot{\gamma}(t_{0},.)=\dot{\gamma}_{0}\in
H_{1}%
\]

\begin{theorem}
The Cauchy problem with the above data for the Einstein equations with $S^{1}
$ isometry group (polarized case) has, if $T-t_{0\text{ }}$is small enough, a
solution with $\gamma\in$ $C^{0}([t_{0},T],H_{2})$ $\cap C^{1}([t_{0}%
,T],H_{1});$ $\lambda,N,\nu\in C^{0}([t_{0},T],$ $W_{3}^{p},1<p<2$ and $N>0$
while $\ \sigma\in C^{1}([t_{0},T],C^{\infty})$ with $\sigma_{t}$ uniformly
equivalent to $\sigma_{0}$. This solution is unique up to $t$ parametrization
of $\tau$ and choice of a cross section of $M_{-1}$ over $T_{eich}.$
\end{theorem}

The proof is by solving alternately elliptic systems for $h,\lambda,N,\nu$ on
each $\Sigma_{t}$, the wave equation for $u$ and the differential system
satisfied by $Q.$ The iteration converges if $T-t_{0}$ is small enough.

We will prove that the solution exists for all $t\geq t_{0}>0,$ and for an
infinite proper time, by obtaining a priori estimates of the various norms,
and of a strictly positive lower bound independent of $T$ for $N.$

\section{Energy estimates.}

We omit the index $t$ when the context is clear. We denote by$\parallel
.\parallel$ and $\parallel.\parallel_{g}$the $L^{2}$ norms in the $\sigma$ and
$g$ metric. We denote by $C_{\sigma}$ numbers depending only on $(\Sigma
,\sigma)$. A lower case index m or M denotes respectively the lower or upper
bound of a scalar function on $\Sigma_{t}$. It follows from the equations
satisfied by $N$ and $\lambda,$ by the maximum principle, that
\begin{equation}
0\leq N_{m}\leq N_{M}\leq2\frac{\partial_{t}\tau}{\tau^{2}}\text{ \ and
\ \ }e^{-2\lambda_{m}}\leq\frac{1}{2}\tau^{2}%
\end{equation}

\subsection{First energy estimate.}

Inspired by the hamiltonian constraint we define the \textbf{energy} by%

\begin{equation}
E(t)\equiv\parallel\gamma^{\prime}\parallel_{g}^{2}+\parallel D\gamma
\parallel_{g}^{2}+\frac{1}{2}\parallel h\parallel_{g}^{2}%
\end{equation}

By integrating the hamiltonian constraint over ($\Sigma_{t,}g)$ and using the
constancy of $\tau$ and the Gauss Bonnet theorem we find, with $\chi$ the
Euler characteristic of $\Sigma,$ that%

\begin{equation}
E(t)=\frac{\tau^{2}}{4}Vol_{g}(\Sigma_{t})+4\pi\chi
\end{equation}
and after some manipulations, that
\begin{equation}
\frac{dE(t)}{dt}=\tau\int_{t}(\frac{1}{2}|h|_{g}^{2}+|\gamma^{\prime}%
|^{2})N\mu_{g}.
\end{equation}

\noindent We see that $E(t)$ is a non increasing function of $t$ if $\tau$ is negative.

\subsection{Second energy estimate.}

We define the \textbf{energy }of\textbf{\ gradient }$\gamma$ by
\begin{equation}
E^{(1)}(t)\equiv\int_{\Sigma_{t}}(J_{0}+J_{1})\mu_{g},\text{ \ \ }J_{1}%
=\mid\Delta_{g}\gamma\mid^{2},\text{ \ \ }J_{0}=\mid D\gamma^{\prime}\mid^{2}%
\end{equation}
After lengthy but straightforward calculation we have found (see [1], indices
raised with $g)$%

\begin{equation}
\frac{dE^{(1)}}{dt}-2\tau E^{(1)}=\tau\int_{\Sigma_{t}}\{NJ_{0}+(N-2)(J_{0}%
+J_{1})\}\mu_{g}+Z
\end{equation}%

\[
Z\equiv2\int_{\Sigma_{t}}\{Nh^{ab}\partial_{a}\gamma^{\prime}\partial
_{b}\gamma^{\prime}+2Nh^{ab}\nabla_{a}\partial_{b}\gamma\Delta_{g}%
\gamma+(\nabla_{b}(\partial^{a}N\partial_{a}\gamma)+\tau\partial_{b}%
N\gamma^{\prime})(\partial^{b}\gamma^{\prime})+
\]%

\begin{equation}
\lbrack(2\partial_{a}Nh^{ac}-4N\partial^{c}\gamma\gamma^{\prime})\partial
_{c}\gamma+2\partial^{a}N\partial_{a}\gamma^{\prime}+\gamma^{\prime}\Delta
_{g}N]\Delta_{g}\gamma\}\mu_{g}%
\end{equation}

We set
\begin{equation}
E(t)=\varepsilon^{2},\text{ \ \ }E^{(1)}(t)=\tau^{-2}\varepsilon_{1}^{2}.
\end{equation}

\section{Elliptic estimates.}

We have shown in [1] that if the sum $\varepsilon^{2}+\varepsilon_{1}^{2}$ is
bounded by a number $c$ - \textbf{hypothesis }$H_{c}$ - the quantities $N, $
,$h,$ $\lambda,$ $\nu$ can be bounded on $\Sigma_{t}$ by using the elliptic
equations they satisfy. In particular, denoting by $\lambda_{M}$ the supremum
of $\lambda$ on $\Sigma_{t},$%
\begin{equation}
\frac{1}{\sqrt{2}}|\tau|e^{\lambda_{M}}\leq1+CC_{\sigma_{t}}(\varepsilon
^{2}+\varepsilon_{1}^{2})
\end{equation}
where $C$ denotes a number depending on $c,$ and $C_{\sigma_{t}}$ a Sobolev
constant of ($\Sigma_{t},\sigma_{t})$

We have also obtained estimates for\textbf{\ }$h,$ and for $N$, namely
\begin{equation}
\parallel h\parallel_{L^{\infty}(g_{t})}\leq CC_{\sigma_{t}}|\tau
|\{\varepsilon+(\varepsilon+\varepsilon_{1})^{2}\}
\end{equation}%
\begin{equation}
0\leq2-N_{m}\leq CC_{\sigma_{t}}(\varepsilon^{2}+\varepsilon\varepsilon
_{1}),\text{ \ \ }||DN||_{L^{\infty}(g_{t})}\leq CC_{\sigma_{t}}%
|\tau|(\varepsilon^{2}+\varepsilon\varepsilon_{1}).
\end{equation}
All these estimates would be sufficient to prove that the energies remain
uniformly bounded for all $t\geq t_{0},$ if small enough initially, if we knew
an uniform (in $t$) bound of the Sobolev constants $C_{\sigma_{t}}.$ We can
obtain such a bound only if the total energy, $\varepsilon^{2}+\varepsilon
_{1}^{2}$ decays when $t$ increases.

\section{Corrected energy estimates.}

The decay we are looking for will be obtained through the introduction of
''corrected energies'' whose $t$ derivatives take advantage of the negative
part of the derivative of the corresponding original energy to give a negative
definite contribution.

\subsection{First corrected energy.}

One defines a \textbf{corrected first energy} by the formula, where $\alpha$
is a positive number:
\begin{equation}
E_{\alpha}(t)=E(t)-2\alpha\tau\int_{\Sigma_{t}}(\gamma-\overset{\_}{\gamma
})\gamma^{\prime}\mu_{g},\text{ \ \ with \ \ }\overset{\_}{\gamma}=\frac
{1}{Vol_{\sigma}\Sigma_{t}}\int_{\Sigma_{t}}\gamma\mu_{\sigma}%
\end{equation}
We estimate the complementary term using the Poincar\'{e} inequality which
gives
\begin{equation}
||\gamma-\overset{\_}{\gamma}||_{g_{t}}\leq e^{\lambda_{M}}||\gamma
-\overset{-}{\gamma}||_{\sigma_{t}}\leq e^{\lambda_{M}}\Lambda_{\sigma_{t}%
}^{-1/2}||D\gamma||_{\sigma_{t}}.
\end{equation}
therefore, by the Cauchy-Schwarz inequality, since $||D\gamma||_{\sigma_{t}%
}=||D\gamma||_{g_{t}}$
\begin{equation}
|\tau\int_{\Sigma_{t}}\gamma^{\prime}(\gamma-\overset{\_}{\gamma})\mu_{g}%
|\leq|\tau|e^{\lambda_{M}}\Lambda_{\sigma_{t}}^{-1/2}||\gamma^{\prime}%
||_{g}||D\gamma||_{g}.
\end{equation}
We deduce from this inequality that:
\[
E(t)\leq\frac{1}{1-a_{t}}E_{\alpha}(t)\text{ \ \ with \ \ }a_{t}\equiv
\frac{\alpha|\tau|e^{\lambda_{M}}}{\Lambda_{\sigma_{t}}^{\frac{1}{2}}}%
\]
We will have $a_{t}<1$ if
\begin{equation}
\alpha<\frac{\Lambda_{\sigma_{t}}^{\frac{1}{2}}}{|\tau|e^{\lambda_{M}}}%
\end{equation}
We have seen that there exists numbers $C$ and $C_{\sigma}$ such that
\begin{equation}
|\tau|e^{\lambda_{M}}\leq\sqrt{2}(1+CC_{\sigma}(\varepsilon^{2}+\varepsilon
_{1}^{2}))
\end{equation}
We suppose that there exist numbers $\Lambda>0$ and $\delta>0,$ independent of
$t,$ such that for all $t$ it holds that
\begin{equation}
\Lambda_{\sigma_{t}}^{\frac{1}{2}}\geq\Lambda^{\frac{1}{2}}(1+\delta)
\end{equation}
then it holds that
\begin{equation}
\frac{\Lambda_{\sigma_{t}}^{\frac{1}{2}}}{|\tau|e^{\lambda_{M}}}\geq
\frac{\Lambda^{\frac{1}{2}}(1+\delta)}{\sqrt{2}(1+CC_{\sigma}(\varepsilon
^{2}+\varepsilon_{1}^{2}))}>\frac{1}{\sqrt{2}}\Lambda^{\frac{1}{2}}%
\end{equation}
as soon as
\begin{equation}
CC_{\sigma}(\varepsilon^{2}+\varepsilon_{1}^{2})<\delta
\end{equation}
When this inequality is satisfied we can choose any number $\alpha$ such that
\begin{equation}
\alpha\leq\frac{1}{\sqrt{2}}\Lambda^{\frac{1}{2}}%
\end{equation}
and so insure that $a_{t}<1.$ For instance if
\begin{equation}
CC_{\sigma}(\varepsilon^{2}+\varepsilon_{1}^{2})<\frac{\delta}{2}%
\end{equation}
then
\begin{equation}
1-a_{t}\equiv1-\frac{\alpha|\tau|e^{\lambda_{M}}}{\Lambda_{\sigma_{t}}%
^{\frac{1}{2}}}\geq\frac{\delta}{2(1+\delta)}%
\end{equation}

\subsection{Decay of the corrected energy.}

We set:%

\[
\frac{dE_{\alpha}}{dt}=\frac{dE}{dt}-R_{\alpha}%
\]
with (the terms explicitly containing the shift $\nu$ give an exact divergence
which integrates to zero)
\[
R_{\alpha}=2\alpha\tau\int_{\Sigma_{t}}\{\partial_{0}\gamma^{\prime}%
(\gamma-\overset{\_}{\gamma})+\gamma^{\prime}\partial_{0}(\gamma-\overset
{\_}{\gamma})-N\tau\gamma^{\prime}(\gamma-\overset{\_}{\gamma})\}\mu_{g}%
\]%

\begin{equation}
+2\alpha\frac{d\tau}{dt}\int_{\Sigma_{t}}\gamma^{\prime}(\gamma-\overset
{\_}{\gamma})\mu_{g}%
\end{equation}
To simplify the writing we suppose that $\int_{\Sigma_{t}}\gamma^{\prime}%
\mu_{g}=0,$ this quantity is conserved in time if $\gamma$ satisfies the wave
equation 3.1. Some elementary computations using 3.1 and integration by parts
show that , using also $\frac{d\tau}{dt}=\tau^{2}:$
\begin{equation}
R_{\alpha}=2\alpha\tau\int_{\Sigma_{t}}\{[|\gamma^{\prime}|^{2}-|D\gamma
|_{g}^{2}]N+\tau(\gamma-\overset{\_}{\gamma})\gamma^{\prime}\}\mu_{g}%
\end{equation}
We write $\frac{dE_{\alpha}}{dt}$ in the form
\begin{equation}
\frac{dE_{\alpha}}{dt}=\tau\int_{\Sigma_{t}}\{2[\frac{1}{2}|h|^{2}%
+(1-2\alpha)|\gamma^{\prime}|^{2}+2\alpha|D\gamma|_{g}^{2}]-2\alpha\tau
\gamma^{\prime}(\gamma-\overset{\_}{\gamma})\}\mu_{g}+\tau A
\end{equation}
where $A$ can be estimated with higher order terms in the energies, using the
inequality 6.3\ satisfied by $N-2,$ since $A$ reads
\begin{equation}
A\equiv\int_{\Sigma_{t}}(N-2)[\frac{1}{2}|h|^{2}+(1-2\alpha)|\gamma^{\prime
}|^{2}+2\alpha|D\gamma|_{g}^{2}]\mu_{g}%
\end{equation}
We look for a positive number $k$ such that the difference $\frac{dE_{\alpha}%
}{dt}-k\tau E_{\alpha}$ can be estimated with higher order terms in the
energies. We have:
\begin{equation}
\frac{dE_{\alpha}}{dt}-k\tau E_{\alpha}=2\tau\int_{\Sigma_{t}}\{[\frac{1}%
{2}|h|^{2}+(1-2\alpha-\frac{k}{2})|\gamma^{\prime}|^{2}+(2\alpha-\frac{k}%
{2})|D\gamma|_{g}^{2}]
\end{equation}%
\begin{equation}
-\alpha(1-k)\tau\gamma^{\prime}(\gamma-\overset{\_}{\gamma})\}\mu_{g}+\tau A
\end{equation}
We have treated in [1] the case where $\Lambda\geq\frac{1}{8},$ $\alpha
=\frac{1}{4}.$ In this case it is possible to take $k=1$ and obtain
immediately
\begin{equation}
\frac{dE_{\frac{1}{4}}}{dt}-\tau E_{\frac{1}{4}}\leq|\tau A|.
\end{equation}
In the general case we have
\begin{equation}
\frac{dE_{\alpha}}{dt}-k\tau E_{\alpha}\leq2\tau\int_{\Sigma_{t}}%
\{[(1-2\alpha-\frac{k}{2})|\gamma^{\prime}|^{2}+(2\alpha-\frac{k}{2}%
)|D\gamma|_{g}^{2}]
\end{equation}%
\begin{equation}
-\alpha(1-k)\tau\gamma^{\prime}(\gamma-\overset{\_}{\gamma})\}\mu_{g}+|\tau A|
\end{equation}
The estimate 7.3 together with the inequality 7.7gives
\begin{equation}
|\tau\int_{\Sigma_{t}}\gamma^{\prime}(\gamma-\overset{\_}{\gamma})\mu_{g}%
|\leq\sqrt{2}\Lambda^{-1/2}||\gamma^{\prime}||_{g}||D\gamma||_{g}%
\end{equation}
Therefore it holds that:
\[
\int_{\Sigma_{t}}\{[(1-2\alpha-\frac{k}{2})|\gamma^{\prime}|^{2}%
+(2\alpha-\frac{k}{2})|D\gamma|_{g}^{2}]-\alpha(1-k)\tau\gamma^{\prime}%
(\gamma-\overset{\_}{\gamma})\}\mu_{g}\geq
\]%
\begin{equation}
(1-2\alpha-\frac{k}{2})||\gamma^{\prime}||_{g}^{2}+(2\alpha-\frac{k}%
{2})||D\gamma||_{g}^{2}-\alpha(1-k)\sqrt{2}\Lambda^{-1/2}||\gamma^{\prime
}||_{g}||D\gamma||_{g}%
\end{equation}
The above quadratic form in $||\gamma^{\prime}||_{g}$ and $||D\gamma||_{g}$ is
non negative if
\begin{equation}
k\leq4\alpha,\text{ \ \ }k\leq2(1-2\alpha)
\end{equation}
and $k$ is such that
\begin{equation}
2\alpha^{2}\Lambda^{-1}(1-k)^{2}-4(2\alpha-\frac{k}{2})(1-2\alpha-\frac{k}%
{2})\leq0
\end{equation}
the inequalities 7.23 imply
\begin{equation}
k\leq1,
\end{equation}

The inequality 7.24 reads%

\begin{equation}
(1-2\Lambda^{-1}\alpha^{2})k^{2}-(1-2\Lambda^{-1}\alpha^{2})2k-2\Lambda
^{-1}\alpha^{2}+8\alpha(1-2\alpha)>0
\end{equation}
That is, since $1-2\Lambda^{-1}\alpha^{2}>0,$%

\begin{equation}
k^{2}-2k+1-\frac{(1-4\alpha)^{2}}{(1-2\Lambda^{-1}\alpha^{2})}>0
\end{equation}
equivalently
\begin{equation}
k<1-\frac{1-4\alpha}{(1-2\Lambda^{-1}\alpha^{2})^{\frac{1}{2}}}%
\end{equation}
There will exist such a $k>0$ if
\begin{equation}
\frac{1-4\alpha}{(1-2\Lambda^{-1}\alpha^{2})^{\frac{1}{2}}}<1
\end{equation}
that is
\begin{equation}
-2\Lambda^{-1}\alpha-16\alpha+8>0.
\end{equation}
i.e.
\begin{equation}
\alpha<\frac{4}{8+\Lambda^{-1}}%
\end{equation}
We have
\begin{equation}
\frac{4}{8+\Lambda^{-1}}\leq\frac{1}{4},\text{ \ \ if \ \ }\Lambda\leq\frac
{1}{8}%
\end{equation}
An elementary computation shows that
\begin{equation}
\frac{4}{8+\Lambda^{-1}}\leq\frac{\Lambda^{\frac{1}{2}}}{\sqrt{2}}%
\end{equation}
with the equality satisfied only when \ \ $\Lambda=\frac{1}{8}.$

We choose $\alpha$ such that it satisfies the inequality 7.31, and then $k>0 $
such that it satisfies 7.28.

\subsection{Corrected second energy.}

We define a \textbf{corrected second energy} $E_{\alpha}^{(1)}$ by the
formula
\[
E_{\alpha}^{(1)}(t)=E^{(1)}(t)+2\alpha^{(1)}\tau\int_{\Sigma_{t}}\Delta
_{g}\gamma\gamma^{\prime}\mu_{g}%
\]
Using again the Cauchy Schwarz inequality, and the Poincar\'{e} inequality to
estimate $||\gamma^{\prime}||_{g_{t}}$ in terms of $||D\gamma^{\prime
}||_{g_{t}}$ (the hypothesis $\overset{\_}{\gamma^{\prime}}=0$ is not
necessary here because on a compact manifold $\int_{\Sigma_{t}}\Delta
_{g}\gamma\mu_{g}=0.)$ we find, (with the same $a_{t}$ as in the previous
subsection)
\begin{equation}
E^{(1)}(t)\leq\frac{1}{1-a_{t}}E_{\alpha}^{(1)}(t)
\end{equation}

\subsection{Decay of the second corrected energy.}

We have found in [1], by straightforward but lengthy computations with the use
of the wave equation for $\gamma$ and $^{(3)}R_{0}^{c}\equiv-N\nabla_{a}%
h^{ac}=2\partial_{0}\gamma\partial^{c}\gamma$ together with Cauchy Schwarz and
Sobolev theorems, an equality of the form:
\[
\frac{dE_{\alpha}^{(1)}}{dt}\equiv\frac{dE^{(1)}}{dt}+R_{\alpha}^{(1)}%
\]
where, with the choice $\tau=\frac{-1}{t}$,
\[
R_{\alpha}^{(1)}\equiv2\alpha^{(1)}\tau\int_{\Sigma_{t}}\{-N|D\gamma^{\prime
}|^{2}+N|\Delta_{g}\gamma|^{2}+(N+1)\tau\Delta_{g}\gamma\gamma^{\prime}%
\}\mu_{g}+\tau Z_{\alpha}%
\]
where $Z_{\alpha},$ given by:
\begin{equation}
Z_{\alpha}\equiv2\alpha\int_{\Sigma_{t}}\partial^{a}N(\partial_{a}\gamma
\Delta_{g}\gamma+\gamma^{\prime}\partial_{a}\gamma^{\prime})+2Nh^{ab}%
\nabla_{a}\partial_{b}\gamma\gamma^{\prime}+2\gamma^{\prime}\partial_{c}%
\gamma(\partial_{a}Nh^{ac}-\gamma^{\prime}\partial^{c}\gamma)\}\mu_{g}%
\end{equation}
can be estimated with higher order terms in the energies. Using 5.6\ we obtain that:%

\[
\frac{dE_{\alpha}^{(1)}}{dt}-(2+k)\tau E_{\alpha}^{(1)}=\tau\int_{\Sigma_{t}%
}\{(2N-2-k-2\alpha N)J_{0}+(2\alpha N+N-2-k)J_{1}\}\mu_{g}%
\]%
\begin{equation}
+2\alpha\tau^{2}\int_{\Sigma_{t}}\{(N+1-2-k)\Delta_{g}\gamma\gamma^{\prime
}\}\mu_{g}+Z+\tau Z_{\alpha}%
\end{equation}
which we write:%

\[
\frac{dE_{\alpha}^{(1)}}{dt}-(2+k)\tau E_{\alpha}^{(1)}=\tau\int_{\Sigma_{t}%
}\{(2-k-4\alpha)J_{0}+(4\alpha-k)J_{1}\}\mu_{g}%
\]%
\begin{equation}
+2\alpha\tau^{2}\int_{\Sigma_{t}}\{(1-k)\Delta_{g}uu^{\prime}\}\mu_{g}+Z+\tau
Z_{\alpha}+Z_{N}%
\end{equation}
where $Z_{N}$ can be estimated with higher order terms in the energies through
the estimate of $N-2.$

The same estimates as those done for the first corrected energy show that,
under the hypothesis made previously, it holds that the term linear in the
energies on the right hand side of the above equality is always non negative
for the following choices of $\alpha$ and $k:$

\begin{itemize}
\item 1. $\Lambda\geq\frac{1}{8}.$ We can choose $\alpha=\frac{1}{4}$ and $k=1.$

\item 2. $\Lambda<\frac{1}{8}.$ We must then choose $\alpha$ and $k$
satisfying the inequalities 7.31 and 7.28.
\end{itemize}

In all cases the estimate of the higher order terms in the energies are the
same ones as obtained in [1], and the following equality holds:
\[
\frac{dE_{\alpha}^{(1)}}{dt}=(2+k)\tau E_{\alpha}^{(1)}+|\tau|^{3}B
\]
where $B$ is a polynomial in first and second derivatives of $\gamma
,h,Dh,DN,D^{2}N$ whose many terms can all be bounded using previous estimates
by a polynomial in $\varepsilon$ and $\varepsilon_{1}$ whose terms are at
least of degree 3 and the coefficients bounded by $CC_{\sigma_{t}},$ under the
$H_{c}$ hypothesis, with $c>0$ a given appropriate number.

\section{Decay of the total energy.}

We define $y(t)$ to be the \textbf{total corrected energy} namely:
\[
y(t)\equiv E_{\alpha}(t)+\tau^{-2}E_{\alpha}^{(1)}%
\]
\bigskip It bounds the total energy $x(t)\equiv E_{tot}(t)\equiv
\varepsilon^{2}+\varepsilon_{1}^{2}$ by
\[
x(t)\equiv\varepsilon^{2}+\varepsilon_{1}^{2}\leq\frac{1}{1-a_{t}}y(t)\text{
\ \ with \ \ }a_{t}\equiv\frac{\alpha|\tau|e^{\lambda_{M}}}{\Lambda
_{\sigma_{t}}^{\frac{1}{2}}}%
\]

We make the following a priori hypothesis, for all $t\geq t_{0}$ for which the
considered quantities exist

\begin{itemize}
\item \textbf{Hypothesis H}$_{\sigma}$ : 1. The numbers $C_{\sigma_{t}}$ are
uniformly bounded by a constant $M_{\sigma}$.
\end{itemize}

2. There exist $\Lambda>0$ and $\delta>0$ such that the inequality 7.6 is satisfied.

\begin{itemize}
\item \textbf{Hypothesis H}$_{E}$. The energies $\varepsilon_{t}^{2}$ and
$\varepsilon_{1,t}^{2}$ satisfy the inequality 7.8.
\end{itemize}

We choose $\alpha$ such that (the case $\Lambda\geq\frac{1}{8},$
\ \ $\alpha=\frac{1}{4}$ was considered in [1]).
\begin{equation}
\alpha<\frac{4}{8+\Lambda^{-1}}<\frac{1}{4}\ \ \text{with}\ \ \Lambda<\frac
{1}{8}%
\end{equation}
Under the hypotheses H$_{c},$H$_{E}$ and H$_{\sigma}$ there exists a number
$M>0$ such that, for all $t:$%
\begin{equation}
1-a_{t}\geq M>0.
\end{equation}

Under the hypothesis H$_{E}$ all powers of $E_{tot}$ greater than 3/2 are
bounded by the product of $E_{tot}^{3/2\text{ }}$ by a constant.

We denote by $M_{i}$ any given number dependent on the bounds of these H's
hypothesis but independent of $t$.

Under the hypotheses H$_{c},$ H$_{\sigma}$ and H$_{E}$ the function $y$
satisfies a differential inequality of the form
\begin{equation}
\frac{dy}{dt}\leq-\frac{k}{t}(y-M_{1}y^{3/2})
\end{equation}
We suppose that $y_{0}\equiv y(t_{0})$ satisfies
\begin{equation}
y_{0}^{1/2}<\frac{1}{2M_{1}}%
\end{equation}
Then $y$ starts decreasing, continues to decrease as long as it exists and
satisfies an inequality which gives by integration, after setting $y=z^{2},$
\[
\log\{\frac{z(1-M_{1}z_{0})}{(1-M_{1}z)z_{0}}\}+\frac{1}{2}k\log\frac{t}%
{t_{0}}\leq0\text{ \ \ a fortiori \ \ }t^{k}y\leq\frac{t_{0}^{k}y_{0}%
}{(1-M_{1}z_{0})^{2}}%
\]
hence, using the hypotheses and previous bounds, the \textbf{decay estimate}
\[
t^{k}x(t)\leq M_{2}x_{0}\text{ \ \ with \ \ }M_{2}\leq\frac{4t_{0}^{k}%
}{(1-a_{t})(1-a_{0})}\leq\frac{4t_{0}^{k}}{M^{2}}%
\]

\section{Teichmuller parameters.}

We require the metric $\sigma_{t}$ to remain, when t varies, in some cross
section of $M_{-1}$ over the Teichmuller space $\mathcal{T}$.

Given a metric s $\in M_{-1}$the \textbf{Dirichlet energy} $D_{s}(\sigma)$ of
the metric $\sigma\in M_{-1}$ is the energy of the (unique) harmonic
diffeomorphism homotopic to the identity $\phi:$($\Sigma,\sigma)\rightarrow
(\Sigma,s).$ It can be written by conformal invariance as
\[
D_{s}(\sigma)\equiv\int_{\Sigma}g^{ab}\partial_{a}\Phi^{A}\partial_{b}\Phi
^{B}s_{AB}(\Phi)\mu_{g}\text{ }%
\]
Let $\sigma_{0}$ satisfy the hypothesis $H_{\sigma}$, then there exists a
number D such that if
$\vert$%
D($\sigma)-D(\sigma_{0})|\leq D,$ called \textbf{Hypothesis }$H_{D},$ then
$\sigma$ satisfies also the hypothesis $H_{\sigma}$. We now estimate $D(\sigma).$

We have if $\Phi$ is a harmonic map that
\[
\frac{d}{dt}D_{s}(\sigma)=\int_{\Sigma_{t}}\{\overset{\_}{\partial}_{0}%
g^{ab}\partial_{a}\Phi^{A}\partial_{b}\Phi^{B}-N\tau g^{ab}\partial_{a}%
\Phi^{A}\partial_{b}\Phi^{B}\}s_{AB}(\Phi)\mu_{g}%
\]
with
\[
\overset{\_}{\partial}_{0}g^{ab}=2Ne^{-4\lambda}h^{ab}+Ne^{-2\lambda}%
h^{ab}\tau
\]
hence
\[
\frac{d}{dt}D_{s}(\sigma)=\int_{\Sigma_{t}}2Ne^{-2\lambda}h^{ab}\partial
_{a}\Phi_{\sigma}^{A}\partial_{b}\Phi_{\sigma}^{B}s_{AB}(\Phi_{\sigma}%
)\mu_{\sigma}%
\]
Using $0<N\leq2$ and e$^{-2\lambda}\leq\frac{\tau^{2}}{2}$ and the bound of
$\parallel h\parallel_{\infty}$ we find:
\[
\frac{d}{dt}D_{s}(\sigma)\leq|\tau|CC_{\sigma}[\varepsilon+(\varepsilon
+\varepsilon_{1})^{2}]D_{s}(\sigma)
\]

Under the hypotheses that we have made the Dirichlet energy satisfies the
differential inequality
\[
\frac{1}{D_{s}(\sigma)}\frac{d}{dt}D_{s}(\sigma)\leq\frac{CM_{\sigma}%
M_{2}^{\frac{1}{2}}x_{0}^{\frac{1}{2}}}{t^{1+\frac{k}{2}}}%
\]
By integration and elementary \ calculus we obtain the inequality, valid for
all $t\geq t_{0}$ since $k$ is a strictly positive number,
\[
|D(\sigma_{t})-D(\sigma_{0})|\leq M_{3}x_{0}^{\frac{1}{2}}%
\]

\section{Global existence.}

\begin{theorem}
Let ($\sigma_{0},q_{0})\in C^{\infty}(\Sigma_{0})$ and ($u_{0},\overset{.}%
{u}_{0})\in H_{2}(\Sigma_{0},\sigma_{0})\times H_{1}(\Sigma_{0},\sigma_{0})$
be initial data for the polarized Einstein equations with U(1) isometry group
on the initial manifold $\Sigma_{0}\times U(1)$ ; suppose that $\sigma_{0}$ is
such that $R(\sigma_{0})=-1.$Then there exists a number $\eta>0$ such that if
\[
x_{0}\equiv E_{tot}(t_{0})<\eta
\]
these Einstein equations have a solution on $\Sigma\times S^{1}\times\lbrack
t_{0},\infty),$ with initial values determined by $\sigma_{0},q_{0}%
,u_{0},\overset{.}{u}_{0}.$ The solution has an infinite proper time extension
since $N_{m}>0.$ It is unique with $\tau=-t^{-1}$ and $\sigma_{t}$ in a chosen
cross section of $M_{-1}$ over $T_{eich}.$
\end{theorem}

Proof. The same continuity argument as in [1].

It can be proved that when $t$ tends to infinity the obtained solution tends
to a metric of the type 1.1.

$^{1}$ Supported in part for the NSF contract n$%
{{}^\circ}%
$ PHY-9732629 and PHY-0098084\ to Yale university

Aknowledgements. We thank L.\ Andersson for suggesting the use of corrected
energies. We thank the university Paris VI, the ITP in Santa Barbara, the
university of the Aegean in Samos, the IHES in Bures and the Schroedinger
Institute in Vienna for their hospitality during our collaboration.

\textbf{References}

[1]\textbf{\ }Y.\ Choquet-Bruhat and V.\ Moncrief, Future global einsteinian
spacetimes with U(1) isometry group. C.R. Acad. Sci. Paris t. 332 serie I
(2001), 137-144. Detailed version in Ann. H. Poincar\'{e} 2 (2001) 1007-1064.

[2] V.\ Moncrief Reduction of Einstein equations for vacuum spacetimes with
U(1) spacelike isometry group, Annals of Physics 167 (1986), 118-142

[3] Y.\ Choquet-Bruhat and V.\ Moncrief Existence theorem for solutions of
Einstein equations with 1 parameter spacelike isometry group, Proc. Symposia
in Pure Math, 59, 1996, H.\ Brezis and I.E.\ Segal ed. 67-80

[4] L.\ Andersson, V.\ Moncrief and A. Tromba On the global evolution problem
in 2+1 gravity J.\ Geom. Phys. 23 1997 n$%
{{}^\circ}%
3-4$,1991-205

\bigskip

{\footnotesize YCB Universit\'{e} Paris 6, 4 Place Jussieu 75232 Paris France
Email YCB@CCR.jussieu.fr}

{\footnotesize VM Department of Physics, Yale University, New Heaven CT, USA
Email vincent.moncrief@yale.edu}
\end{document}